\documentstyle[multicol,aps,epsf]{revtex}
\def\d{{\rm d}}
\begin{document}
\title{The `Friction' of Vacuum, and other Fluctuation--Induced Forces}
\author{Mehran Kardar}
\address{Department of Physics, Massachusetts Institute of  
Technology, Cambridge, MA 02139}
\author{Ramin Golestanian}
\address{Institute for Advanced Studies in Basic Sciences,
Zanjan   45195-159, Iran}
\date{\today}
\maketitle
\begin{abstract}

The {\it static} Casimir effect describes an attractive force between two conducting plates, due
to {\it quantum fluctuations} of the electromagnetic (EM) field in the intervening space.
{\it Thermal fluctuations} of correlated fluids (such as critical mixtures, super-fluids,
liquid crystals, or electrolytes) are also modified by the boundaries, resulting in finite-size
corrections at criticality, and additional forces that effect wetting and layering 
phenomena. Modified fluctuations of the  EM field can also account for the `van der
Waals' interaction between conducting spheres, and have analogs in the
fluctuation--induced interactions between inclusions on a membrane.
We employ a path integral formalism to study these phenomena for boundaries
of arbitrary shape. This allows us to examine the many unexpected phenomena
of the {\it dynamic} Casimir effect due to moving boundaries.
With the inclusion of quantum fluctuations, the EM vacuum behaves essentially 
as a complex fluid, and modifies the motion of objects through it. 
In particular, from the mechanical response function of the EM vacuum, we
extract a plethora of interesting results, the most notable being:
{\bf (i)} The effective mass of a plate depends on its shape, and becomes anisotropic.
{\bf (ii)} There is dissipation and damping of the motion, again dependent upon shape
and direction of motion, due to emission of photons.
{\bf (iii)} There is a continuous spectrum of resonant cavity modes that can be
excited by the motion of the (neutral) boundaries.    

\end{abstract}
\begin{multicols}{2}

\section{Outline}

Fluctuation-induced forces are ubiquitous in nature, covering many topics 
from biophysics to cosmology\cite{Casimir,Dzyaloshinskii,Mostepanenko,Krech,Weinberg}.
There are two basic ingredients in these phenomena:
{\bf (i)} A fluctuating medium, such as the electromagnetic (EM) field; and 
{\bf (ii)} External objects whose presence suppresses (or in some way modifies) 
the fluctuations, such as dipoles or conductors.
The overall strength of the interaction is proportional to the driving energy of   
fluctuations ($k_B T$ and $\hbar$ for thermal and quantum fluctuations,  respectively);
its range is related to that of the correlations of the fluctuations.
The most interesting cases are when the interactions are long--ranged,
corresponding to scale free fluctuations.

The goal of this article is to provide a glimpse of the unity and simplicity of
fluctuation--induced forces.
While we attempt to describe a wide range of phenomena, this selection is
by no means exhaustive, and highly biased by subjective interests.
In the spirit of a colloquium, we have tried to avoid technical details, 
preferring to present dimensional arguments whenever possible.
The interested reader is referred to various sources for calculational details.

A prototype of fluctuation--induced interactions is the Casimir force between
conducting plates, due to {\it quantum fluctuations} of the EM field.  
In Sec.\ref{secFIF} we discuss several cases were the source of interaction 
is  the {\it thermal fluctuations} of a correlated fluid between the bounding plates.
This interaction was originally proposed for a liquid mixture at its critical
point, but is also present when long-range correlations appear as a
result of symmetry breaking, as in {\it superfluids} or {\it liquid crystals}.
There is even an attractive component to the (mainly repulsive) force between 
two similarly charged plates, due to fluctuations of counterions in a neutralizing solution. 
Since the latter connection is seldom made explicit, 
we expand on its origin in Appendix \ref{ChargeA}.
While the reader may skip anyone of these sections without losing general track of 
the article, we note that experiments on wetting of helium films may provide
a beautiful test of these forces.

The van der Waals and London dispersion forces between atoms and molecules
can also be attributed to the modified fluctuations of the EM field.
As we point out in Sec.\ref{secVDW}, there are analogous forces
between inclusions on a membrane.
Their origin is the modified {\it surface} fluctuations, and they decay more slowly 
with separation than the standard van der Waals interaction.
We use this example to emphasize that {\it non-additivity} is an  important feature 
of fluctuation--induced forces: 
they cannot be obtained from a pairwise sum of two-body potentials.

Many new results are obtained in going beyond the simple geometries 
of flat plates and simple spheres, by looking at rough and deformed
structures, as in Sec.\ref{secROUGH}.
Our key to implementing the corresponding non-standard boundary 
conditions is a path integral approach, 
which is sketched in Appendix \ref{roughA}.
This approach can also be used in conjunction with a 
path integral quantization of the electromagnetic field.
Since in this relativistic theory, deformations in space and time appear
on the same footing, we can examine the {\it dynamic Casimir} effect
which is introduced in Sec.\ref{secDC}.

Some of the unexpected phenomena that emerge from 
quantum fluctuations of the EM field in the presence of moving 
deformed plates are discussed in Sec.\ref{secCM}.
There are corrections to the mass of a plate that depend on its shape.
There is also dissipation due to emission of photons (which accounts 
for the `friction' in the title of this article).
While these effects are typically very small, we believe that they
are significant for what they imply about the nature of the quantized EM vacuum.
Qualitatively, with the inclusion of quantum fluctuation, the vacuum behaves
as a complex fluid which influences the bodies moving through it.

\section{Fluctuation--Induced Forces}\label{secFIF}

\subsection{Quantum fluctuations}

The standard Casimir effect\cite{Casimir,Mostepanenko} is a macroscopic  
manifestation of quantum fluctuations of vacuum. 
In 1948, Casimir considered the electromagnetic field in the  cavity formed by
two conducting plates at a separation $H$. 
Because the electric field must vanish at the boundaries, the normal modes of the
cavity are characterized by wave-vectors
${\vec k}=\left(k_x, k_y, {\pi n/ H}\right)$, with integer $n$.
Once quantized, these normal modes can be regarded as harmonic
oscillators of frequencies $\omega(\vec k)=c|\vec k\,|$; each of which in its
ground state has energy $\hbar\omega(\vec k)/2$. 
While adding up all the ground state energies leads to an infinite contribution
to the overall energy ${\cal E}(H)$, Casimir showed that a finite {\it attractive} 
force is obtained from
\begin{equation}\label{CasF}
F(H)=-{\partial {\cal E}\over\partial H}=-\hbar c \times{A\over H^4} \times {\pi^2\over240},
\end{equation}
where $A$ is the area of the plates. Thus, by measuring the mechanical 
force between macroscopic bodies, it is in principle possible to
gain information about the behavior of the quantum vacuum.

The predictions of Casimir where followed by experiments on quartz\cite{AD}
and aluminum\cite{Spar} plates at separations $H>10^3\AA$. However,
these experiments, and others reviewed in Ref.\cite{CasX}, provided results
that were at best in qualitative agreement with Eq.(\ref{CasF}). It is only
quite recently that high precision measurements of the force (using a torsion
pendulum) between a gold plate, and gold plated sphere, confirmed the
accuracy of the theoretical prediction to within \%5\cite{LamX}.

\subsection{Thermal fluctuations}

While the Casimir interaction is due to the {\it quantum fluctuations} of the
electromagnetic field, there are several examples in {\it classical statistical
mechanics}, where forces are induced by the {\it thermal fluctuations}  of
a correlated fluid. One of the best known examples comes from the finite
size corrections to the free energy at a critical point\cite{Krech}. 
Fisher and de Gennes\cite{FG} argued that in a binary liquid mixture,
the concentrations near a wall are perturbed only over a distance of the order
of the correlation length $\xi$. Any interaction mediated by the concentration
fluctuations must also decay with this characteristic length.
However, at the critical point. where $\xi$ diverges, they suggested
an attractive contribution to the free energy of a critical film,
that varies with its thickness $H$, as
\begin{equation}\label{FdG}
\delta {\cal F}(H)= -k_BT\times {A\over H^2}\times \Delta .
\end{equation}
This is to be expected on dimensional grounds, as the free energy comes
from thermal fluctuations, hence proportional to $k_BT$, 
and must be extensive in $A$.
(Similar analysis in $d$-dimensions leads to a dependence as $1/H^{d-1}$.)
In two dimensions, exact values for the dimensionless amplitude $\Delta$
can be obtained by employing techniques of conformal field
theories\cite{BCN}. In higher dimensions, they can be estimated 
numerically\cite{NI}, and by $\epsilon=4-d$ expansions\cite{KD}. 

In analogy to the Casimir energy, we can regard Eq.(\ref{FdG})
as due to the modified free energy of concentration fluctuations
by the boundaries.
However, the force that results from this free energy decays as $1/H^3$.
The difference in power of $H$ from Eq.(\ref{CasF}) is explained by noting that
the fluctuation energy in the latter is quantum in origin, hence
proportional to $\hbar c$, which has dimensions of {\it energy
times length}.  

\subsection{Superfluid films}

In fact, long-range forces are induced by thermal fluctuations of any
{\it correlated medium}, by which we mean  any system with fluctuations
that have long-range correlations \cite{phonons}. 
The critical system is a very particular example; 
much more common are cases where long-range correlations
exist due to Goldstone modes of a broken continuous symmetry, as
in superfluids or liquid crystals. A superfluid is characterized by a complex
order parameter, whose phase $\phi$ may vary across the system. The 
energy cost of such variations is governed by the Hamiltonian
\begin{equation}\label{superH}
{\cal H}[{\phi]}={K\over 2}\int \d^3 {\bf x}\left(\nabla \phi\right)^2,
\end{equation}
where the ``phonon stiffness" $K$ is related to the superfluid density.
There is no characteristic length scale for fluctuations of $\phi$,
which scale as a power of the observation length. Consequently,
we expect power law finite size scaling, just as in the case of a
critical point.
In the Casimir geometry, the free energy  resulting from
thermal fluctuations of these modes has the form\cite{LiK}
\begin{equation}\label{superF}
\delta {\cal F}(H)= -k_BT\times {A\over H^2}\times {\zeta(3)\over 16\pi}.
\end{equation}
Note that the result is universal, i.e. independent of the stiffness $K$.
A similar expression is obtained for the free energy of the
electromagnetic field confined between metallic plates
at high temperatures $k_BT\gg \hbar c/H$. However, the result 
is larger by a factor of two\cite{SDM}, reflecting
the two polarizations of the normal modes (photons).

Liquid Helium is a powerful wetting agent, 
which tends to spread over most surfaces. The thickness of the
wetting layer is controlled by the strength of the attractive
forces that bind the film to the substrate\cite{Dietrich},
mostly due to van der Waals interactions. In the presence of
a chemical potential penalty of $\delta \mu$ per unit volume, the energy of a film 
of thickness $H$ is
\begin{equation}\label{HeE}
{E(H) \over k_B T}=A\left[ \frac{\delta\mu}{k_B T} H+  {C\over H^2}\right],
\end{equation}
where $C$ is a {\it positive} numerical constant. 
Minimizing this expression leads to a thickness \cite{Retardation}
\begin{equation}\label{HeH}
H=\left({2 C k_B T \over \delta\mu}\right)^{1/3}.
\end{equation}

When the helium film is in the normal phase, 
the film thickness is determined solely by the
strength of the van der Waals (vdW) force.  
The numerical value of $C_{>}=C_{vdW} >0$ 
depends on the substrate, and is nonuniversal.
However, when the film becomes superfluid, there is an additional 
attractive fluctuation--induced (FI) force due to Eq.(\ref{superF}), and
$C_{<}=C_{vdW} +C_{FI}$;  where $C_{FI}=-\zeta (3)/16\pi \approx -0.02391$.
In the vicinity of the superfluid transition, there is a different attractive contribution
to the force due to finite size scaling (FSS) of the critical fluctuations, as in Eq.(\ref{FdG}),
and $C_{\lambda}=C_{vdW} +C_{FSS}$. The best estimate for the finite size scaling
amplitude  for the XY model in $d=3$ is $C_{FSS}  \approx -0.03$ \cite{Indek2,Mon}.
The parameter $C$ thus takes three different values in the normal fluid,
at the $\lambda$-point, and in the superfluid phase.
From Eq.(\ref{HeH}) we then expect {\it two} jumps in the film thickness, 
as the temperature is lowered through the superfluid transition.
Experiments to monitor the film thickness, thus providing a test of these forces,
are currently underway at Penn. State University\cite{Moses}. 

\subsection{Liquid crystals}

Liquid crystals exemplify anisotropic cases of correlated fluids due to 
broken symmetry, which again lead to fluctuation--induced forces\cite{Mik,Adj,LiK}. 
They are also easily accessible, as experiments can be performed at room
temperature and require no fine tuning to achieve criticality. 
A {\it nematic} liquid crystal is composed of long molecules that are aligned,
with an order parameter which is the `director' field  ${\bf  n}( {\bf r})$, 
characterizing the local preferred direction of the 
long axis of the molecules\cite{rDG}. 
The energy cost of fluctuations of this field is 
given by\cite{rDG}
\begin{eqnarray}\label{nematic}
{\cal H}_N={1\over 2}\int \d^3 {\bf r} &[& \kappa_1 ({\bf \nabla}\cdot
{\bf n})^2+\kappa_2 ({\bf n}\cdot{\bf \nabla} \times {\bf n})^2\nonumber\\
&+&\kappa_3 ({\bf n}\times {\bf \nabla} \times {\bf n})^2 ] .
\end{eqnarray}
\noindent
Integrating over the nematic fluctuations leads to a free energy contribution
\begin{equation}\label{Fnem}
\delta {\cal E}_N=-k_BT\times {A\over H^2}\times {\zeta(3)\over16\pi}\left(
             {\kappa_3\over \kappa_1}+{\kappa_3\over \kappa_2}\right).
\end{equation}
Note that the resulting force does depend on the relative strengths
of the elastic coupling constants (reflecting the anisotropy of the system).

In a smectic liquid crystal, the molecules segregate into layers which 
are  fluid like. The fluctuations of these layers from perfect stacking
are described by a scalar deformation $u({\bf x}, z)$, which 
is subject to a Hamiltonian
\begin{equation}\label{smectic}
{\cal H}_S={1\over 2}\int \d^3 {\bf r}\left[B\left({\partial u\over
\partial z}\right)^2+\kappa\left(\nabla^2 u\right)^2\right].
\end{equation}
The resulting interaction energy
\begin{equation}\label{Fsme}
\delta {\cal E}_S=-k_BT\times {A\over H\lambda}\times {\zeta(2)\over16\pi},
\quad{\rm with}\quad\lambda\equiv\sqrt{\kappa\over B},
\end{equation}
falls off as $1/H$, reflecting the extreme anisotropy which has introduced
an additional length scale $\lambda$ into the problem.
For potential experimental tests of these forces in surface freezing
of liquid crystal films, see Ref.\cite{Lyra}.

\subsection{Charged fluids}

Interactions between a collection of {\it charged macroions} in an aqueous 
solution of  neutralizing {\it counterions}, with or without added salt, are in 
general very complex. The macroions may be charged spherical 
colloidal particles, charged amphiphilic membranes, stiff 
polyelectrolytes (e.g. microtubules, actin filaments, and DNA), or 
flexible polyelectrolytes (e.g. polystyrene sulphonate), and the 
counterions could be mono- or polyvalent. 
It is known that, under certain conditions, the accumulation (condensation) 
of counterions around highly charged macroions can turn the
repulsive Coulomb interaction between them into an 
attractive one.  The attractive interaction is induced by the 
diminished charge-fluctuations close to the macroions, 
(due to the condensation of counterions)\cite{Oosawa,Marcelja,Attard},
and in this sense related to the effects discussed in the previous sections.

Since the connection between the entropic attraction of charged macroions
and the general class of fluctuation--induced forces, is seldom made explicit,
in Appendix \ref{ChargeA} 
we present a path integral formulation 
that makes this analogy more transparent. 
The interaction between macroions can be broken into two parts: 
A Poisson-Boltzmann (PB) free energy, and a fluctuation-induced correction.
Specifically, consider two parallel negatively charged 2D plates 
with densities $-\sigma$, separated by a distance $H$ in $d=3$, 
in a solution of neutralizing counterions with valence $z$. 
The PB equation can be solved exactly in this geometry, and the 
corresponding PB free energy, in the limit of highly charged plates, is\cite{FPB}
\begin{equation} \label{FPB1}
F_{PB}={\pi \over 2} \times \frac{A}{z^2 \ell_B H} \; \left[1+
\frac{1}{4 \pi^2 \ell_B^2 z^2 \sigma^2 H^2}+\cdots \right],
\end{equation} 
in which $\ell_B\equiv e^2/\epsilon k_B T$ is the {\it Bjerrum length}.
Note that in the limit $\ell_B z \sigma H \gg 1$, the interaction is independent
of the charge densities of the plates; i.e. it is {\it universal}. 

The fluctuation-induced correction involves calculation of a 
determinant (see Appendix \ref{ChargeA}), which depends on
the local charge compressibilities. 
The true compressibility profile (and the charge density
profile) emerging from of the solution of the PB equation,
is generally very complicated. 
It is usual to simplify the problem by
assuming that the surface charge density is so high that the counterions
are confined to a layer of thickness $\lambda \ll H$, where
 $\lambda \sim 1/z \ell_B  \sigma$ is the Gouy-Chapman length. 
Then we can use an approximate compressibility profile 
$m^2(z)=(2/\xi)[\delta(z+H/2)+\delta(z-H/2)]$, 
in which $\xi^{-1} =\pi^2 \ell_B^2 z^2 \sigma^2 \lambda$
defines a ``crossover length". In the limit  $H/\xi \gg 1$, we again obtain \cite{Att2}
\begin{equation}   \label{FFIGC}
F_{FI}= -k_BT\times {A\over H^2}\times {\zeta(3)\over 16\pi}
\;\left[1+O(\xi/H)\right],
\end{equation}
for the fluctuation-induced part of the interaction\cite{ChargeA}.

\section{Dispersion Forces}\label{secVDW}

\subsection{Van der Waals interactions}

In addition to his work on the force between plates, Casimir also
realized\cite{CasPol} that the van der Waals and London\cite{London} 
forces can also be understood on the same footing:
The presence of the atoms modifies the fluctuations of the
electromagnetic field, resulting in an attractive interaction.
(For a modern perspective, see the discussion by Kleppner in Ref.\cite{Kleppner}.)

For example, let us consider two conducting spheres of volumes
$V_1=4\pi a_1^3/3$ and $V_2=4\pi a_2^3/3$, at a distance $R$.
Usually, the van der Waals interaction is obtained from the instantaneous
induced dipoles on the spheres. However, it can be equivalently obtained
by examining the electromagnetic fluctuations of the remaining space.
The fluctuation induced interaction is proportional to the product of the
excluded volumes, and thus on dimensional grounds we expect a potential
\begin{equation}\label{vdWT}
{\cal V}(R)=-k_BT\times {V_1V_2\over R^6}\times \Delta_T,
\end{equation}
due to thermal fluctuations. 
When the fluctuations are of quantum origin, Eq.(\ref{vdWT}) is modified to
\begin{equation}\label{vdWQ}
{\cal V}(R)=-\hbar c\times {V_1V_2\over R^7}\times \Delta_Q,
\end{equation}
with $\Delta_Q=23/(4\pi)$\cite{CasPol}. 

Let us compare the above result with the more standard approach to
calculating the London force between two neutral {\it atoms}:
While the average dipole for each atom is zero, an instantaneous
dipole fluctuation in one can induce a parallel instantaneous dipole 
on the other, leading to an attraction. Since the direct dipole--dipole
interaction decays as $1/R^3$, the induced effect scales as the
square, i.e.  $1/R^6$. In this regards, it is similar to the result in
Eq.(\ref{vdWT}), except that the characteristic energy is set by
a typical atomic excitation energy of $\hbar\omega_0$ rather than $k_BT$.
The retardation effects are then obtained by taking  into account the 
finite speed of light. We can imagine that for large enough distances,
when the signal goes from atom-1 to atom-2 to induce the dipole,
and return to atom-1 to induce an attraction, it  finds the dipole at atom-1 
somewhat misaligned; resulting in a weaker attraction. 
The characteristic time for electron movements can be estimated from 
the frequency of the orbit as $\tau=2\pi/\omega_0$.
The crossover occurs when the travel time for the signal is 
comparable to this characteristic time, namely $R/c \sim \tau$.
Hence, taking into account the retardation effect, the interaction is 
\begin{equation}\label{vdWR}
{\cal V}_{r}(R)=-\hbar\omega_0\times {V_1V_2\over R^6}
\times f\left({R \omega_0\over c}\right).
\end{equation}
The crossover function $f(x)$ is a constant for $x \to 0$, and vanishes as
$1/x$ for $x \to \infty$. In the latter limit, the dependence on $\omega_0$ 
vanishes, and the Casimir--Polder result of Eq.(\ref{vdWQ}) is recovered. 
A similar crossover function between the two forms of interaction for 
conducting spheres in Eqs.(\ref{vdWT}) and (\ref{vdWQ}),
occurs at a distance $\lambda_T\sim \hbar c/k_BT$.

\subsection{Inclusions on membranes}

Dispersion forces are not limited to particles in three dimensional space,
but also occur for inclusions on films and membranes, the latter is of
potential importance for understanding the interactions between proteins
floating on a cell membrane.  A membrane is a bilayer of
{\it amphiphilic} molecules, each composed of a {\it hydrophilic} or polar
head, and a {\it hydrophobic} tail of hydrocarbon chains. 
The polar heads prefer to be in contact with the water, and in the bilayer
structure insulate the ``oily" hydrocarbon interior from contact with water.
A bilayer that is in equilibrium with amphiphiles in solution, 
can easily change its area by exchanging molecules with this reservoir. 
This  implies that the surface tension is zero \cite{DGT,BL,DL},
and the energy cost of deforming the bilayer is entirely due to bending\cite{CH}.
The ``flicker" of the membrane is thus governed by the elastic Hamiltonian
\begin{equation} \label{CHH}
{\cal H}=\frac{\kappa}{2} \int \d ^2 {\bf x} (\nabla^2 h( {\bf x}))^2, 
\end{equation} 
where $h( {\bf x})$ is a height function describing deformations of the surface.
Typical values of the bending rigidity $\kappa$, of biological membranes
is of the order of $10^2\,^\circ K$.

Cell membranes also include proteins that perform various biological
functions (e.g. pumps).
Each protein inclusion disturbs the lipid bilayer, resulting in interactions
between nearby inclusions (c.f. Refs.\cite{isr,mou,goul,dan} and
references therein). 
These disturbances, and the resulting interactions, tend to be short-ranged, 
falling off exponentially with a characteristic length related to the distance 
over which the lipid membrane ``heals''\cite{dan}. 
There are also longer ranged interactions between the proteins: 
In addition to the standard van der Waals interaction, there are interactions
mediated by the disturbed fluctuations of the flickering membrane.
As discussed in Ref.\cite{goul}, such interactions exist as long as the rigidity 
of the inclusion differs from that of the ambient membrane, and fall off as $1/R^4$.
In particular, if the inclusions are much stiffer than the membrane, 
the fluctuation--induced potential is
\begin{equation}\label{}
{\cal V}(R)=-k_BT\times{A^2\over R^4}\times{6 \over \pi^2 },
\end{equation}
where $A$ is the area of each inclusion\cite{correct}.
The interaction is  attractive and independent of $\kappa$ and 
$\bar\kappa$; its energy scale set by $k_BT$.
A generalization of this result that includes quantum fluctuations of
membranes is given in Ref.\cite{HSK}.

\begin{figure}
\epsfysize=2.0truein
\centerline{\epsffile{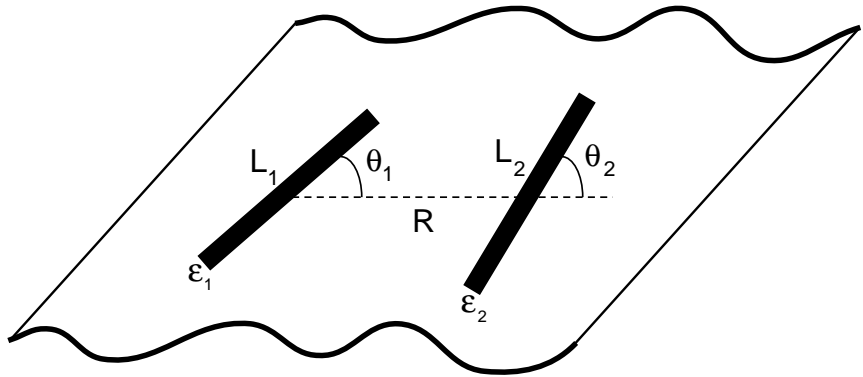}}
\bigskip
\caption{Two rod-shaped inclusions embedded in a membrane. The rods are separated
by a distance $R$. The $i$th rod has length $L_i$, width $\epsilon_i$, and
makes an angle $\theta_i$ with the line joining the centers of the two rods.}
\label{rods}
\end{figure}

The form of the interaction depends sensitively on the shapes of the
inclusions, as demonstrated by the calculation of the fluctuation--induced
interaction  between rod-like objects\cite{GGK}.
The rods are assumed to be sufficiently rigid so that they do not deform
coherently with the underlying membrane. They can thus only perform
rigid translations and rotations while remaining attached to the surface.
As a result, the fluctuations of the membrane are constrained,
having to vanish at the boundaries of the rods.
Consider the situation depicted in Fig.~\ref{rods}, with  two rods of lengths 
$L_1$ and $L_2$ at a separation $R\gg L_i$. 
The fluctuation-induced interaction is given by
\end{multicols}
\begin{equation}\label{memb}
V^{T}(R,\theta_1,\theta_2)=-\frac{k_{B}T}{128}\times
\frac{L_1^2 L_2^2}{R^4}\times\cos^2\left[2\left(\theta_{1}+\theta_{2}
\right)\right]+O\left({1\over R^6}\right),
\end{equation}
\bigskip
\begin{multicols}{2}\noindent
where $\theta_1$ and $\theta_2$ are the angles between the
rods and the line adjoining their centers, as indicated in Fig.~\ref{rods}.

The orientational dependence is the {\it square}  of a quadrupole--quadrupole
interaction, with the unusual property of being minimized for both 
parallel and perpendicular orientations of the rods.
The above fluctuation-induced interactions decay less rapidly at
large distances than van der Waals forces and may play an important 
role in aligning asymmetric inclusions in biomembranes. Since orientational
correlations are often easier to measure than forces, this result may
also be useful as a probe of the fluctuation-induced interaction.
Finally, this interaction could give rise to novel two-dimensional
structures for collections of rodlike molecules. In particular, the
resemblance of the orientational part of the interaction to dipolar forces
suggests that a suitable way to minimize the energy of a collection of rods 
is to form them into chains. (If the rods are not colinear, the interactions
cannot be simultaneously minimized.) 

An important property of fluctuation-induced interactions is that they are 
{\it non--additive}, and cannot be obtained by adding two-body potentials.
For example, consider an interaction $U(|r_1-r_2|)du_1du_2$, 
between any two infinitesimal segments of two rods in Fig.~\ref{rods}.
If both rods are of length $L$ at a distance $R\gg L$, expanding $ |r_1-r_2|$, 
and integrating over the two rods, leads to the interaction
\begin{eqnarray}	\label{add}
V(R,\theta_1,\theta_2)&=&L^2U(R)+{L^4 \over 6}\left( {U'(R) \over R}
+U''(R)\right) \\
&-&\frac{L^4}{12}\left(  {U'(R) \over R}
-U''(R)\right)(\cos2\theta_{1}+\cos2\theta_{2}).\nonumber
\end{eqnarray}
The angular dependence is now completely different, and
minimized when the two rods are parallel to their axis of separation.
Presumably both interactions are present for rods of finite thickness;
the additive interaction is proportional to $L^2(L\epsilon/R)^2$, where
$\epsilon$ is the thickness. The previously calculated interactions
are thus larger by a factor proportional to $(R/\epsilon)^2$ and
should dominate at large separations.

The case of stiff linear inclusions at close separations ($L\gg R$)
is considered in Ref.\cite{RG}. It is shown that a finite
rigidity of the linear inclusions leads to a screening out of 
the Casimir-type fluctuation-induced attraction. The screening 
length is set by the ratio between the rigidity of the polymer
and that of the membrane. This is the length scale below which
the polymers are seen as straight parallel lines, hence resulting
in a Casimir interaction. Moreover,
the attractive interactions could lead to an instability in the shape
of the stiff polymers signalling a major reduction in their rigidity
(softening) induced by the membrane fluctuations\cite{RG}. 

\section{Rough surfaces}\label{secROUGH}

Most computations of Casimir forces are for simple geometries, e.g. between
two parallel plates, or perfect spheres. It is natural to consider how these forces are
modified by the roughness that is present in most ``random" surfaces.
There are a number of calculations that go beyond the simple planar
geometry. For example, in Ref.\cite{BalDup} a multiple scattering approach
is used to compute the interactions for arbitrary geometry in a perturbation 
series in the curvature. 
A generalization of the approach due to Dzyaloshinskii, Lifshitz,
and Pitaevskii \cite{Dzyaloshinskii} is developed in Ref.\cite{Novikov}
to study the Casimir forces for surfaces with roughness.
A phenomenological approach is introduced in Refs.\cite{Bordag} in
which small deviations from plane parallel geometry are treated by
using an additive summation of Casimir potentials. However (as demonstrated
in the previous section), fluctuation induced forces are not additive,
and additional steps are necessary to correct the result\cite{Bordag}.
Another perturbative approach is introduced in Ref.\cite{Ford}, which 
could in principle be used to treat surfaces with roughness, although
not explicitly carried out in this paper. Most of these approaches
suffer from rather cumbersome treatments of the boundary conditions.

In Ref.\cite{LiK}, a path integral approach is introduced that makes possible
relatively simple computations of the fluctuation induced force.
(A sketch of this formulation is presented in Appendix \ref{roughA}.) 
This approach has a number of advantages. First, different manifolds 
(with arbitrary intrinsic and embedding dimensions) in various correlated 
fluids can be treated in a similar fashion. Second, the boundary conditions 
are quite easily implemented, and the corrections can be
computed perturbatively in the deformations. 
While this method was originally developed for the study of thermal
fluctuations, it can be adapted to quantum fluctuations, as discussed
in the next section. In the remainder of this section we calculate the
corrections to the thermal Casimir force due to substrate roughness. 

Many solid surfaces produced by rapid growth or
deposition processes are characterized by self--similar fluctuations\cite{KX}. 
The fluctuations of a self--affine surface grow as
\begin{equation}\label{Rou}
\overline{[h({\bf x})-h({\bf y})]^2}=A_S  |{\bf x}-{\bf y}|^{2\zeta_S}\ ,
\end{equation}
where the overbar denotes quenched average, and $\zeta_S$ is a characteristic
roughness exponent. 
The Casimir force between a flat and a such a rough surface (with a
correlated fluid in between) is calculated in Ref.\cite{LiK}. 
The resulting free energy per unit area is
\begin{equation}\label{Casi}
{\cal F}(H)=-\frac{k_B T}{H^2}\frac{\zeta (3)}{16\pi}
-\frac{ k_B T A_S L^{2\zeta_S}}{H^4}\frac{3\zeta(3)}{16\pi}
+\frac{k_B T A_S}{H^{4-2\zeta_S}}\frac{C_1}{4}\ ,
\end{equation}
where  $C_1$ is a numerical coefficient \cite{LiK}, and
$L$ is the extent (upper cutoff) of the self-affine structure, satisfying
$\Delta H \equiv A_S^{\frac{1}{2}}L^{\zeta_S}\ll H$. (This is the condition
that the total width due to roughness, $\Delta H$ is less than the average
separation $H$, so that the plates are not in contact.)
As long as $L\gg H \gg \Delta H$, the interactions
in Eq.(\ref{Casi}) are arranged in order of increasing strength. The largest
effect of randomness is to increase the Casimir attraction by an amount 
proportional to $(\Delta H/H)^2$. There is also another correction term,
of the opposite sign, that decays as $1/H^{4-2\zeta_S}$, and in principle
can be used to indirectly measure the roughness exponent $\zeta_S$.
In Eq.(\ref{Rou}), if all lengths are measured in units of an atomic scale $a_0$ 
(e.g. the diameter of a surface atom), $A_S$ becomes dimensionless. Using 
a reasonable set of parameters: $\zeta_S\approx 0.35$, $a_0\approx 5 {\rm \AA}$, 
$A_S \approx 1$ and $L\approx 300 {\rm \AA}$, we estimate that for surfaces of 
$1 {\rm mm}$ size, and $100 {\rm \AA}$ apart, the forces generated by the three terms
in Eq.(\ref{Casi}) are $1.9\times 10^{-4}$,  $4.9\times 10^{-5}$, and
$3.7\times 10^{-6}{\rm N}$ respectively, (using a
reasonable lower cutoff of $\sim 20 {\rm \AA}$),
which is measurable with the current force  apparatus\cite{LamX}.

\section{The Dynamic Casimir Effect}\label{secDC}

\subsection{Background}

Although less well known than its static counterpart, the
dynamical Casimir effect, describing the force and radiation
from moving mirrors has also garnered much attention
\cite{Moore,Fulling,Jaekel,Neto,Cavity,Meplan,Lambrecht}.
This is partly due to connections to Hawking and Unruh effects
(radiation from black holes and accelerating bodies, respectively),
suggesting a deeper link between quantum mechanics,
relativity, and cosmology\cite{Davis,Weinberg}.

The creation of photons by moving mirrors was first obtained 
by Moore\cite{Moore} for a 1 dimensional cavity.  
Fulling and Davis\cite{Fulling} demonstrated that there is a
corresponding  force even for a single mirror,
which depends on the third time derivative of its displacement.
These computations take advantage of conformal symmetries of the
1+1 dimensional space time, and can not be easily generalized to 
higher dimensions. Furthermore, the calculated force has causality 
problems reminiscent of the radiation reaction forces in classical 
electron theory \cite{Jaekel}. It has been shown that this problem 
is an artifact of the unphysical assumption of perfect reflectivity
of the mirror, and can be resolved by considering realistic 
frequency dependent reflection and transmission from the 
mirrors \cite{Jaekel}. 

Another approach to the problem starts with the fluctuations in the
force on a single plate. The fluctuation--dissipation theorem is then
used to obtain the mechanical response function\cite{Neto},
whose imaginary part is related to the dissipation. This method
does not have any causality problems, and can also be extended
to higher dimensions. (The force in 1+3 dimensional space-time
depends on the fifth power of the motional frequency.)
The emission of photons by a perfect cavity, and the observability
of this energy, has been studied by different approaches
\cite{Cavity,Meplan,Lambrecht}. The most promising candidate is
the  resonant production of photons when the mirrors
vibrate at the optical resonance frequency of the cavity\cite{Davis}. 
A review, and more extensive references are found in Ref.\cite{Barton}.
More recently, the radiation due to vacuum fluctuations of a 
collapsing bubble has been proposed\cite{Schwinger,Eberlein,Knight} as a possible 
explanation for the intriguing phenomenon of sonoluminescense.
(Subsequent experimental measurements of the duration of the
signal\cite{sonoX} may favor more classical explanations.)

A number of authors have further discussed to notion of {\it frictional forces}:
Using conformal methods in 1+1 dimensions, Ref. \cite{Dodonov1} finds
a friction term
\begin{equation} \label{frictionforce}
F_{\rm friction}(H)= \alpha \;F_{\rm static}(H)\;\left({\dot{H} /
c}\right)^{2},
\end{equation}
for slowly moving boundaries, where $\alpha$ is a numerical constant 
that only depends on dimensionality.
The additional factor of $(v/c)^2$ makes detection of
such a force yet more delicate.
There are a few attempts to calculate forces (in higher dimensions) 
for walls that move {\it laterally}, i.e. parallel to each other
\cite{Levitov,Mkrt,Barton2,Pendry,EberleinPW}: 
It is found that {\it boundaries that are not ideal conductors} experience a
friction as if the plates are moving in a viscous fluid. 
This friction has a complicated dependence
on the frequency dependent resistivity of the plates, and vanishes
in the cases of ideal (nondissipating) conductors or dielectrics.
The ``dissipation'' mechanism for this ``friction'' is by inducing eddy currents 
in the nonideal conductors, and thus distinct from the Casimir effect.
Possible experimental evidence of such a contribution to friction has been 
recently reported in Ref. \cite{Dayo}.
The experiment employs a quartz crystal microbalance technique to measure 
the friction associated with sliding of solid nitrogen along a lead 
surface, above and below the superconducting transition temperature of lead. 
An abrupt drop in friction is reported  at the transition point as the substrate 
enters the superconducting state \cite{Dayo}. 

An interesting analog of the dynamic
Casimir effect is suggested for the moving interface between two different
vacuum states of superfluid 3He \cite{Volovik}. In this system, the
Andreev reflection of the massless ``relativistic" fermions
which live on the A-phase of the interface provides
the corresponding mechanism for friction: The interface is analogous
to a perfectly reflecting wall moving in the quantum vacuum.   

\subsection{Path integral formulation}

The path integral methods originally developed for rough surfaces\cite{GKPRL}
can also be applied to the  problem of perfectly reflecting 
mirrors that undergo arbitrary dynamic deformations. 
Consider the path integral quantization of a scalar field $\phi$ with
the action
\begin{equation}
S=\frac{1}{2} \int \d^4 X \;\partial_{\mu}\phi(X)
			\partial_{\mu}\phi(X),\label{action}
\end{equation}
where summation over $\mu=0,\cdots, 3$ is implicit.  
Following a Wick rotation, imaginary time appears as another
coordinate $X_4=ict$, in the $4$-dimensional space-time. 
In principle, we should use the
electromagnetic vector potential $A_\mu(X)$, but requirements of
gauge fixing complicate the calculations, while the final results only
change by a numerical prefactor. (We have explicitly
reproduced the known result for gauge fields between flat plates 
by this method\cite{GKPRL}.)
We would like to quantize the field subject to the constraints 
of its vanishing on a set of $n$ manifolds (objects) defined
by $X=X_{\alpha}(y_{\alpha})$, where $y_{\alpha}$ parametrize 
the $\alpha$th manifold. We implement the constraints using delta
functions, and write the partition function as
\begin{eqnarray}
{\cal Z}&=&\int {\cal D}\phi(X)
		\prod_{\alpha=1}^{n} \prod_{y_{\alpha}}
		\delta\left(\phi\left(X_{\alpha}(y_{\alpha})
		\right)\right)\;\exp\left\{-\frac{S[\phi]}{\hbar} 
		\right\}.\label{Z1} 
\end{eqnarray}

The delta functions are next represented by integrals over Lagrange  
multiplier fields. Performing the Gaussian integrations over $\phi(X)$ 
then leads to an effective action for the Lagrange multipliers 
which is again Gaussian \cite{LiK}. Evaluating ${\cal Z}$ is thus 
reduced  to calculating the logarithm of the determinant of a kernel. 
Since the Lagrange multipliers are defined on a set of manifolds
with nontrivial geometry, this calculation is generally complicated. 
To be specific, we focus on  two parallel 2d plates 
embedded in 3+1 space-time, and separated by an 
average distance $H$ along the $x_3$-direction. Deformations of the
plates are parametrized by the height functions $h_1({\bf x},t)$ and  
$h_2({\bf x},t)$, where ${\bf x}\equiv(x_1,x_2)$ denotes the two lateral 
space coordinates, while $t$ is the time variable. 
As sketched in Appendix \ref{roughA},
$\ln {\cal Z}$ can be calculated by a perturbative series in
powers of the height functions \cite{h/H}. The resulting expression for the 
effective action (after rotating back to real time), defined by 
$S_{\rm eff}\equiv -i \hbar \ln {\cal Z}$, and eliminating $h$
independent terms, is
\end{multicols}
\begin{eqnarray}
S_{\rm eff}=\frac{\hbar c}{2} \int\frac{\d \omega \d^2 {\bf  
q}}{(2\pi)^3}\;
\left[A_{+}(q,\omega)\left(|h_1({\bf q},\omega)|^2+|h_2({\bf  
q},\omega)|^2\right) 
\hskip 5 cm \right. \label{Seff} \\ 
\left. \hskip 2 cm -A_{-}(q,\omega)\left(h_1({\bf q},\omega)
h_2(-{\bf q},-\omega)+h_1(-{\bf q},-\omega)h_2({\bf q},\omega)\right)  
\right]
+O(h^3).\nonumber
\end{eqnarray}
\begin{multicols}{2}

\subsection{The response function}

The kernels $A_{\pm}(q,\omega)$, that are closely related to the  
mechanical response of the system (see below), are functions of 
the separation $H$, but depend on ${\bf q}$ and $\omega$ only 
through the combination $Q^2=q^2-\omega^2/c^2$. The closed 
forms for these kernels involve cumbersome integrals, and are 
not very illuminating.  Instead of exhibiting these formulas, 
we shall describe their behavior in various regions of the 
parameter space. In the limit 
$H \rightarrow \infty$, $A_{-}^{\infty}(q,\omega)=0$, and
\begin{equation}
A_{+}^{\infty}(q,\omega)=\frac{1}{360 \pi^2 c^5}\left\{
	\begin{array}{cl}
	- (c^2 q^2-\omega^2)^{5/2}
	& \mbox{for $\omega < c q$} , \\
	 \\
	i\,\,{\rm sgn}(\omega) (\omega^2- c^2  
q^2)^{5/2}
	& \mbox{for $\omega > c q$}, 
	\end{array} \right. \label{A+infty}
\end{equation}
where ${\rm sgn}(\omega)$ is the sign function. 
While the effective action is real for $Q^2>0$, it becomes purely
imaginary for $Q^2<0$.
The latter signifies dissipation of energy \cite{Neto}, presumably  
by generation of photons \cite{Lambrecht}.  It agrees precisely 
with the results obtained previously\cite{Neto} for the special 
case of flat mirrors (${\bf q}=0$).
(Note that dissipation is already present for a single mirror.)

In the presence of a second plate (i.e. for finite $H$), the 
parameter space of the kernels subdivides into three different 
regions as depicted in Fig.~2. In region I ($Q^2>0$ for any $H$), 
the kernels are finite and real, and hence there is no dissipation.
In region IIa where $-\pi^2/H^2 \leq Q^2 <0$, the 
$H$-independent part of $A_{+}$ is imaginary, while the 
$H$-dependent parts of both kernels are real and finite.
(This is also the case at the boundary $Q^2=-\pi^2/H^2$.)
The dissipation in this regime is simply the sum of what would 
have been observed if the individual plates were decoupled, 
and unrelated to the separation $H$. 
By contrast, in region IIb where $Q^2 < -\pi^2/H^2$, both 
kernels diverge with infinite real and imaginary parts\cite{cutoff}.
This $H$-dependent divergence extends all the way to the negative
$Q^2$ axis, where it is switched off by a $1/H^5$ prefactor.
\begin{figure}
\epsfysize=2.8truein
\centerline{\epsffile{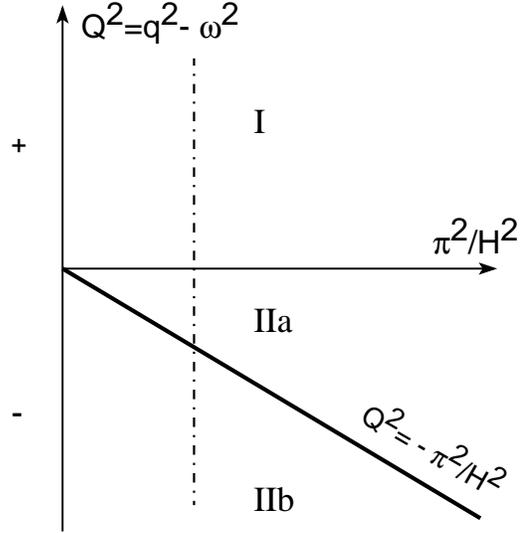}}
\bigskip
\centerline{\caption{Different regions of the $({\bf q},\omega)$ plane.}}
\label{phase}
\end{figure}

As a concrete example, let us examine the lateral vibration of  
plates with fixed roughness, such as two corrugated mirrors.
The motion of the plates enters through the time dependencies 
$h_1({\bf x},t)=h_1({\bf x}-{\bf r}(t))$ and 
$h_2({\bf x},t)=h_2({\bf x})$; i.e. the first plate undergoes lateral  
motion described by ${\bf r}(t)$, while the second plate is stationary.
The lateral force exerted on the first plate is obtained from 
$f_i(t)=\delta S_{\rm eff}/\delta r_i(t)$. Within linear response, 
it is given by 
\begin{equation}
f_i(\omega)=
\chi_{ij}(\omega)\;r_j(\omega)+f_i^{0}(\omega),\label{rough1}
\end{equation}
where the ``mechanical response tensor'' is 
\end{multicols}
\begin{equation}
\chi_{ij}(\omega)=\hbar c \int \frac{\d^2 q}{(2\pi)^2}
\;q_i q_j \left\{\left[A_{+}(q,\omega)-A_{+}(q,0)\right]\;|h_1({\bf  
q})|^2 
+\frac{1}{2} A_{-}(q,0) \left(h_1({\bf q})h_2(-{\bf q})+h_1(-{\bf q})
h_2({\bf q})\right) \right\}, \label{rough}
\end{equation}
and there is a residual force
\begin{equation}
f_i^{0}(\omega)=-\frac{\hbar c}{2}\;2\pi\delta(\omega)\; 
\int \frac{\d^2 q}{(2\pi)^2}
\;i q_i A_{-}(q,0) \left(h_1({\bf q})h_2(-{\bf q})
-h_1(-{\bf q})h_2({\bf q})\right). \label{fi0w}
\end{equation}
\begin{multicols}{2}

\section{Corrugated Mirrors}\label{secCM}

Let us now consider a corrugated plate with a deformation 
$h_1({\bf x})=d \cos({{\bf k}\cdot {\bf x}})$.
From the frequency--wavevector dependence of the mechanical 
response function in Eq.(\ref{rough}), we extract a plethora of interesting
results, some of which we discuss next.

\subsection{Mass corrections}
For a single plate ($H\to\infty$), we can easily calculate
the response tensor using the explicit formulas in Eq.(\ref{A+infty}). 
In the limit of $\omega\ll ck$, expanding the result in powers of  
$\omega$ gives $\chi_{ij}=\delta m_{ij}\omega^2+O(\omega^4)$,
where $\delta m_{ij}=A\hbar d^2 k^3 k_i   k_j/(288\pi^2 c)$,
can be regarded as corrections to the mass of the plate.
(Cut-off dependent mass corrections also appear, as in Ref.\cite{Barton}.) 
Note that these mass corrections are {\it anisotropic} with
\begin{eqnarray}\label{deltaM}
\delta m_{\parallel}&=&\frac{1}{288\pi^2 }\frac{\hbar }{c}A  d^2 k^5,\nonumber\\
\delta m_{\perp}&=&0.
\end{eqnarray}
Parallel and perpendicular components are defined with respect to
${\bf k}$, and $A$ denotes the area of the plates. 
The mass correction is inherently very small: For a macroscopic  
sample with $d\approx \lambda=2\pi/k \approx 1$mm, density 
$\approx 15 {\rm gr/cm}^3$, and thickness $t\approx 1$mm, we find 
$\delta m/m \sim 10^{-34}$. Even for deformations of a microscopic  
sample of atomic dimensions (close to the limits of the applicability 
of our continuum representations of the boundaries),  $\delta m/m$
can only be reduced to around $10^{-10}$. 

With the second plate at a separation $H$,  the mass renormalization
becomes a function of both $k$ and $H$, with a crossover from the
single plate behavior for $k H \sim 1$. In the limit of $k H \ll 1$,
we obtain $\delta m_{\parallel}=\hbar A B k^2 d^2/48 c H^3$ and 
$\delta m_{\perp}=0$, with $B=-0.453$. Compared to the single plate,
there is an enhancement by a factor of $(kH)^{-3}$ in $\delta m_\parallel$.
While the actual changes in mass are immeasurably small, the
hope is that its {\it anisotropy} may be more accessible, say by
comparing oscillation frequencies of a plate in two orthogonal directions.

\subsection{Dissipation:}
For $\omega\gg ck$ the response function is imaginary, and we define  
a frequency dependent effective shear viscosity by 
$\chi_{ij}(\omega)=-i\omega\eta_{ij}(\omega)$. This viscosity is also
anisotropic, with
\begin{eqnarray}\label{deltaN}
\eta_{\parallel}(\omega)&=&{1\over 720\pi^2}{\hbar\over  c^4} A 
d^2k^2  \omega^4,\nonumber\\
\eta_{\perp}(\omega)&=&0.
\end{eqnarray}
Note that the dissipation is proportional to the fifth time derivative 
of displacement, and there is no dissipation for a uniformly 
accelerating plate. However, a freely oscillating plate will undergo 
a damping of its motion. The characteristic decay time for a plate of
mass $M$ is $\tau\approx 2M/\eta$. For the macroscopic plate of 
the previous paragraph, vibrating at a frequency of 
$\omega\approx 2ck$ (in the $10^{12}$Hz range), the decay time is  
enormous, $\tau\sim 10^{18}$s. However, since the decay time 
scales as the fifth power of the dimension, it can be reduced to 
$10^{-12}$s, for plates of order of 10 atoms. However, the required 
frequencies in this case (in the $10^{18}$Hz range) are very large.
Also note that for the linearized forms to remain valid in this high
frequency regime, we must require very small amplitudes, 
so that the typical velocities involved $v\sim r_0\omega$, are 
smaller than the speed of light. The effective dissipation in region
IIa of Fig.~2 is simply the sum of those due to individual plates,
and contains no $H$ dependence.

\subsection{Resonant Emission:}
The cavity formed between the two plates supports a continuous
spectrum of normal modes for frequencies $\omega^2> c^2(k^2+\pi^2/H^2)$.
We find that both real and imaginary parts of $A_{\pm}(k,\omega)$, 
diverge in this regime, which we interpret as resonant dissipation
due to excitation of photons in the cavity.
Resonant dissipation has profound consequences for motion of plates. 
It implies that due to quantum fluctuations of vacuum, 
{\it components of motion with frequencies in the range of 
divergences cannot be generated by any finite external force}! 
The imaginary parts of the kernels are proportional to the total 
number of excited photons \cite{Lambrecht}. Exciting these degrees
of motion must be accompanied by the generation of an infinite 
number of photons, requiring an infinite amount of  energy, and thus
impossible. However, as pointed out in Ref.\cite{Lambrecht}, the 
divergence is rounded off by assuming finite reflectivity and 
transmissivity for the mirrors. Hence, in practice, the restriction is 
softened and controlled by the degree of ideality of  the mirrors in 
the frequency region of interest. 
 
Related effects have been reported in the literature for 1+1 
dimensions\cite{Cavity,Meplan,Lambrecht,Davis}, but occuring at 
a {\it discrete} set of frequencies $\omega_n=n \pi c/H$, with 
integer $n\geq2$. These resonances occur when the frequency of the
external perturbation matches the natural normal modes of the cavity,
thus exciting quanta of such modes. In one space dimension,
such modes are characterized by a discrete set of wavevectors
that are integer multiples of $\pi/H$. The restriction to $n\geq2$ is a
consequence of quantum  electrodynamics being a `free' theory 
(quadratic action): only two-photon states can be excited subject to
conservation of energy. Thus the sum of the frequencies of the two 
photons should add up to the external frequency\cite{Lambrecht}. 
In higher dimensions, the appropriate parameter is the combination 
$\omega^2/c^2-q^2$. From the perspective of the excited photons,  
conservation of momentum requires that their two momenta add up 
to $q$, while energy conservation restricts the sum of their frequencies
to $\omega$. The in-plane momentum $q$, introduces a continuous
degree of freedom: the resonance condition can now be satisfied for a
continuous spectrum, in analogy with optical resonators. 
In Ref.\cite{Lambrecht}, the lowest resonance frequency is found to 
be $2\pi c/H$ which seems to contradict our prediction. However, 
the absence of $\omega_1=\pi c/H$ in 1+1 D is due to a vanishing
prefactor\cite{Lambrecht}, which is also present in our calculations.
However, in exploring the continuous frequency spectrum in higher
dimensions, this single point is easily bypassed, and there is a 
divergence for all frequencies satisfying
$\omega^2/c^2 > q^2+\pi^2/H^2$, where the inequality holds in its  
strict sense. 

\subsection{Radiation spectra}
Where does the energy go when the plates experience viscous dissipation?
When the viscosity is a result of losses in the dielectric 
boundaries\cite{Levitov,Mkrt,Barton2,Pendry,EberleinPW}, the
energy is used up in heating the plates.
Since we have examined perfect mirrors, the dissipated energy can
only be accounted for by the emission of photons into the cavity.
The path integral methods can be further exploited to calculate the
spectrum of the emitted radiation\cite{MGK}. 
The basic idea is to relate the transition amplitude from an empty
vacuum (at $t\to-\infty$) to a state with two photons (at $t\to+\infty$),
to two point correlation functions of the field, 
which is then calculated perturbatively in the deformations.
From the transition amplitude (after integrating over the states of one
photon) we obtain the probability that an emitted photon  is
observed at a frequency $\Omega$, and a particular orientation.

Specifically, calculations of the angular distribution and spectrum 
of radiation were performed\cite{MGK} for
a single perfectly reflecting plate, undulating harmonically with a frequency
$\omega_0$, and a wavevector ${\bf k}_0$. 
Depending on the ratio $\omega_0/k_0$, we find that 
radiation at a frequency $\Omega$ is restricted to a particular window
in solid angle . 
The total spectrum of radiation is found by integrating the angular
distribution over the unit sphere, and is a symmetric function
with respect to $\omega_0/2$, where it is peaked. 
The peak sharpens as the parameter $\omega_0/k_0\to0$, and 
saturates for $k_0=0$.

The connection between the dissipative dynamic Casimir force, and
radiation of photons, is made explicit by calculating the total number of
photons radiated per unit time and per unit area of the plate.
The result is identical to the energy dissipation rate calculated in Ref.\cite{GKPRL}.
No radiation is observed at frequencies higher than $\omega_0$, 
due to conservation of energy, and also for $\omega_0/k_0<1$,
in agreement with section B above, where no dissipative forces 
are found in this regime.

\acknowledgements
We have benefitted from collaborations on these problems with 
M. Goulian, H. Li,  M. Lyra, and F. Miri.  MK is supported by the NSF 
grant DMR-93-03667. RG acknowledges many helpful discussions 
with J. Indekeu, and T. Liverpool, and support from the Institute for 
Advanced Studies in Basic Sciences, Gava Zang, Zanjan, Iran.  

\appendix

\section{Path-Integral Formulation and Deformed Surfaces} \label{roughA}

In this Appendix, we sketch the path integral formulation developed
in Ref.\cite{LiK} for surfaces with roughness.
Consider $n$ manifolds embedded in a $d$-dimensional correlated fluid,
with an energy cost appropriately generalized from Eq.(\ref{superH}).
The manifolds are described by the functions $R_\alpha (x_\alpha)$,
where $x_\alpha$ is a $D_\alpha$-dimensional internal coordinate 
($D_\alpha=1$ for a polymer, and $D_\alpha=2$ for a membrane), while 
$R_\alpha$ indicates a position in the $d$-dimensional fluid.
The fluctuation-induced interactions between the manifolds are obtained
by integrating over all configurations of the field $\phi$,
subject to the constraints of its vanishing on the external manifolds.
The boundary conditions 
($\phi(R_{\alpha}(x_\alpha))=0$, for $\alpha=1, 2, \ldots, n$)
are imposed by inserting delta functions. Using the integral 
representation of the delta function,  we can write 
\end{multicols}
\begin{equation}\label{Seffb}
\exp\left(-\frac{{\cal H}_{eff}}{k_B T}\right)=\frac{1}{{\cal Z}_0}
\int {\cal D}\phi(r)\prod_{\alpha=1}^n{\cal D} \psi_{\alpha}(x_\alpha)
\exp\left(-{\cal H}_0[\phi]+i\int \d x_\alpha \psi_{\alpha}(x_\alpha)
\phi(R_{\alpha}(x_\alpha))\right)\ ,
\end{equation}
where $\psi_{\alpha}(x_\alpha)$ are the auxiliary fields defined on the $n$  
manifolds, acting as sources coupled to $\phi$. For the quadratic Hamiltonian
of Eq.(\ref{superH}), it is easy to integrate over the field $\phi$, and obtain the
long-range interactions between the sources as
\begin{equation}\label{Seffc}
\exp\left(-\frac{{\cal H}_{eff}}{k_B T}\right)=\int\prod_{\alpha=1}^n
{\cal D} \psi_{\alpha}(x_\alpha)\exp 
\left(-{\cal H}_1[\psi_{\alpha}( x_\alpha)]\right)\ .
\end{equation}
The action ${\cal H}_1[\psi_{\alpha}(x_\alpha)]$ is a quadratic form for the 
$n$ component field $\Psi\equiv (\psi_1, \psi_2, \ldots, \psi_n)$, given by
\begin{equation}\label{Snew}
{\cal H}_1[\Psi]\equiv\Psi M\Psi^T=\sum_{\alpha=1}^n\sum_{\beta=1}^n
\int \d x_\alpha \d y_\beta \psi_{\alpha}(x_\alpha)G^d\bigl(R_{\alpha}
(x_\alpha)-R_{\beta}(y_\beta)\bigr)\psi_{\beta}(y_\beta)\ ,
\end{equation}
\begin{multicols}{2}\noindent
where $G^d(r)\equiv\left\langle\phi(r)\phi(0)\right
\rangle_0$ is the two-point correlation function of the field $\phi$ in 
free space. Finally, the effective interaction between the manifolds is
obtained by performing the Gaussian integrations over the field $\Psi$ as 
\begin{equation}\label{S}
{\cal H}_{eff}[R_\alpha (x_\alpha)]=\frac{k_B T}{2}\ln {\det} 
(M[R_\alpha (x_\alpha)])\ .
\end{equation}

The matrix $M$ (which can be read off from Eq.(\ref{Snew})) is a  
functional of $R_{\alpha}(x_\alpha)$ and its 
determinant is in general difficult to evaluate for arbitrary configurations.
It is possible, however, to perturbatively calculate the corrections due to
small deformations around simple base configurations. Consider two surfaces
in $d=3$, with average separation $H$. One plate has small deformations
described by $h({\bf x})$, i.e. $R_1({\bf x})=(x_1,x_2,0)$, while
$R_2({\bf x})=(x_1,x_2,H+h({\bf x}))$. The effective Hamiltonian
can now be written as ${\cal H}_{eff}={\cal H}_{flat}+{\cal H}_{corr}$,
where ${\cal H}_{flat}$ is the Casimir interaction for two flat plates,
and ${\cal H}_{corr}$ is the additional cost of deformations. For flat
plates of area $A$, the interaction energy is
\begin{equation}\label{Sf}
{{\cal H}_{flat}\over A}=k_B T \int {\d^2{\bf p}\over (2\pi)^2}\ln\left(\frac{1}{2\alpha p}\right)-
        \frac{\zeta(3)}{16\pi}\frac{k_B T}{H^2}\ .
\end{equation}
The first term in Eq.(\ref{Sf}) is a contribution to the surface tension which
depends on a lattice cutoff.  The second term is the usual Casimir interaction, 
decaying as $1/H^2$, and with a universal amplitude $-\zeta(3)/16\pi
\approx -0.02391$. 

The energy cost of deformations is given by
\end{multicols}
\begin{eqnarray}\label{Scor}
{\cal H}_{corr}&=&-k_B T\times\frac{3 \zeta (3) }{16\pi H^4} 
\int \d^2{\bf x} h^2({\bf x}) +\frac{k_B T}{4}
         \int \d^2{\bf x} \d^2{\bf y}[h({\bf x})-h({\bf y})]^2\times\nonumber\\
&&\left\{\frac
         {1}{8\pi^2|{\bf x}-{\bf y}|^6}
        +\frac{1}{2\pi|{\bf x}-{\bf y}|^3 H^3}
         K_1\left(\frac{|{\bf x}-{\bf y}|}{H}\right)+
\frac{1}{H^6}\left[K_1\left(\frac
         {|{\bf x}-{\bf y}|}{H}\right)\right]^2+\frac{1}{H^6}\left[K_2\left(\frac
         {|{\bf x}-{\bf y}|}{H}\right)\right]^2\right\}\ ,
\end{eqnarray}
\begin{multicols}{2}\noindent
with two kernel functions  defined by 
\begin{eqnarray}\label{Ker}
K_1(t)&\equiv\int_0^\infty  \frac{\d u}{2\pi }u^2 (e^{2u} -1)^{-1}J_0(tu) ,\\
K_2(t)&\equiv\int_0^\infty \frac{\d u}{2\pi } {u^2 e^u}(e^{2u} -1)^{-1}J_0(tu).
\end{eqnarray}
The first term in Eq.(\ref{Scor}) represents an instability to deformations
that is related to the attraction between the plates. Remarkably, this term 
can be obtained intuitively by replacing $1/{H^2}$ with $1/(H+h({\bf x}))^2$ 
in Eq.(\ref{Sf}) and averaging over the position $\bf x$. The second term
represents long-range interactions between the deformations, induced by the
fluctuations of the field. The 
first term in the curly brackets is the conformation energy of the deformed 
surface in the absence of the second plane, and is independent of $H$.  The 
remaining terms represent correlations due to the presence of second plane. 
Both $K_1(t)$ and $K_2(t)$ approach a constant as $t\rightarrow 0$. As 
$t\rightarrow\infty$, $K_1(t)\sim 1/t^3$, while $K_2(t)\sim \exp(-bt)$, with 
$b\approx 3.3$. The large $t$ behaviors of $K_1(t)$ and $K_2(t)$  determine 
the long--range interactions between height fluctuations.

\section{Path integral formulation of charged fluids}  \label{ChargeA} 

Here, we  introduce a systematic path integral formulation to study
fluctuation-induced interactions in a charged fluid. 
Consider $n$ charged manifolds embedded in a $d$--dimensional 
aqueous solution of neutralizing counterions, interacting through
Coulomb potentials.
The manifolds have charge densities $-\sigma_{\alpha}$ (all assumed
to be negatively charged for simplicity), and are 
described by the functions $R_\alpha (x_\alpha)$,
where $x_\alpha$ is a $D_\alpha$--dimensional internal coordinate,
 while  $R_\alpha$ indicates a position in the $d$--dimensional 
solution. There are  $N_c$ positively charged counterions of valence $z$, 
each described by a position vector $R_i$, in the $d$-dimensional solution. 
The Coulomb Hamiltonian can be written as
\begin{equation}  \label{HC}
H_{C}={1 \over 2} \int \d^d X \d^d X' \; \rho(X) \frac{e^2}
{\epsilon |X-X'|^{d-2}} \rho(X'), 
\end{equation}
where
\begin{eqnarray}  \label{rhoX}
\rho(X)=&-&\sum_{\alpha=1}^{n} \int \d x_\alpha
\sigma_{\alpha}  \delta^d (X-R_\alpha (x_\alpha)) \\
&+&\sum_{i=1}^{N_c} z \delta^d (X-R_i), \nonumber
\end{eqnarray}
is the number density of the charges.
Charge neutrality requires $-\sum_{\alpha=1}^{n} \sigma_\alpha
A_\alpha +z N_c=0$, where $A_\alpha$ is the 
$D_\alpha$-dimensional area of the $\alpha$th manifold.

A restricted partition function of the Coulomb system, 
depending  upon the shapes and locations of the macorions,  
is now given by
\begin{equation}  \label{ZN1}
{\cal Z}_{N_c} [R_\alpha (x_\alpha)]=\int \prod_{i=1}^{N_c} {\d^d R_i \over a^d}
\; e^{- {H_C} /{k_B T}},
\end{equation}
in which $a$ is a short-distance cut-off. Using the Hubbard-Stratanovich 
transformation of the Coulomb interaction,
\end{multicols}
\begin{equation}  \label{Hu-St}
e^{- {H_C}/ {k_B T}}=\int {\cal D} \phi(X) \exp\left\{
-\frac{\epsilon k_B T}{2 S_d e^2} \int \d^d X (\nabla \phi)^2
+ i \int \d^d X \rho(X) \phi(X)\right\},
\end{equation}
we can rewrite the partition function as
\begin{equation}  \label{ZN2}
{\cal Z}_{N_c}[R_\alpha (x_\alpha)]= 
\int {\cal D} \phi(X) \exp\left\{-\frac{\epsilon k_B T}{2 S_d e^2} 
\int \d^d X (\nabla \phi)^2 -i \sum_{\alpha=1}^{n} \int \d x_\alpha \;
\sigma_{\alpha}  \phi(R_\alpha (x_\alpha))\right\} \;
\left(\int  {\d^d R \over a^d} \;e^{i z \phi(R)} \right)^{N_c},
\end{equation}
where $S_d$ is the area of the $d$-dimensional unit sphere. We can 
introduce a fugacity $y$, and a rescaled partition function
\begin{equation}  \label{ZR1}
{\cal Z} [R_\alpha (x_\alpha)]=\frac{y^{N_c}}{N_c !} 
{\cal Z}_{N_c} [R_\alpha (x_\alpha)],
\end{equation}
that can be rewritten as
\begin{eqnarray} \label{ZR2}
{\cal Z} &=& \sum_{N=0}^{\infty} \delta_{N,N_c} 
\frac{y^{N}}{N !}  {\cal Z}_{N} [R_\alpha (x_\alpha)] \\
&=& \sum_{N=0}^{\infty} \int_{0}^{2 \pi} {\d \theta  \over 2 \pi}
e^{i  \theta (N_c-N)}
\int {\cal D} \phi(X) \exp\left\{-\frac{\epsilon k_B T}{2 S_d e^2} 
\int \d^d X (\nabla \phi)^2 -i \sum_{\alpha=1}^{n} \int \d x_\alpha \;
\sigma_{\alpha}  \phi(R_\alpha (x_\alpha))\right\} 
{1 \over N !} \left(y \int  {\d^d R \over a^d} \;e^{i z \phi(R)} \right)^N. \nonumber
\end{eqnarray}
A shift in the field $\phi$ by $-\theta$, and use of the neutrality
condition renders the $\theta$-integration trivial. We can then sum
up the exponential series, and obtain
\begin{equation}  \label{ZR3}
{\cal Z}[R_\alpha (x_\alpha)]= \int {\cal D} \phi(X) \;
e^{-{\cal H}[\phi]}, 
\end{equation}
in which
\begin{equation}  \label{HPhi}
{\cal H}[\phi]=\frac{\epsilon k_B T}{2 S_d e^2} 
\int \d^d X (\nabla \phi)^2 +i \sum_{\alpha=1}^{n} \int \d x_\alpha \;
\sigma_{\alpha}  \phi(R_\alpha (x_\alpha))-{y \over a^d}
\int \d^d X e^{i z \phi(X)}.
\end{equation}
Note that the fugacity $y$ can be eliminated using the identity
\begin{equation} \label{yNc}
N_c=y \frac{\partial \ln {\cal Z}}{\partial y},
\end{equation}
which follows from Eq.(\ref{ZR1}).

We next evaluate the path integral using a saddle point approximation. The extremum
of Eq.(\ref{ZR3}), obtained from $\delta {\cal H}/\delta \phi=0$, is the solution of the Poisson-Boltzmann (PB) equation 
\begin{equation}   \label{PB1}
-\nabla^2(z \psi(X))-\kappa^2 e^{- z \psi(X)}=-\sum_{\alpha=1}^{n} \int \d x_\alpha
\frac{S_d e^2 z \sigma_{\alpha}}{\epsilon k_B T}  \delta^d (X-R_\alpha (x_\alpha)),
\end{equation}
for the (real) field $\psi(X)=-i {\bar \phi}(X)$, in which 
$\kappa^2=S_d e^2 y z^2/\epsilon k_B T a^d$ defines the inverse square of 
the Debye screening length.
To study the fluctuations on top of this saddle point, we can set 
$\phi={\bar \phi}+\delta \phi$, and expand the Hamiltonian up to quadratic
order, to get
\begin{equation} \label{HPhi2}
{\cal H}[\phi]={\cal H}[{\bar \phi}]+\frac{\epsilon k_B T}{2 S_d e^2} 
\int \d^d X \left[(\nabla \delta \phi)^2+\kappa^2 e^{- z \psi(X)} \delta \phi^2 \right].
\end{equation}  
\begin{multicols}{2}
The free energy of the system of charged manifolds in 
the presence of fluctuating counterions now reads
\begin{equation}  \label{F1}
F=F_{PB}+\frac{k_B T}{2} \ln \det \left[-\nabla^2+m^2(X)\right],
\end{equation}
where $F_{PB}={\cal H}[i \psi(X)]$ is the Poisson-Boltzmann free energy,
and  $m^2(X)=\kappa^2 e^{- z \psi(X)}$ is a ``mass (or charge compressibility) profile". 
The PB free energy is known to be generically repulsive
\cite{isr,Oosawa}.
The fluctuation-induced correction, however, is attractive. For highly
charged manifolds, it is indeed
reminiscent of the Casimir interactions, but with the boundary constraints
smoothed out. To see this, one should note that the mass profile is indeed
identical to the density profile of the counterions. Highly charged
manifolds accumulate counterions in their vicinity, and consequently
the fluctuations of the ``potential" field $\phi$ are suppressed in a region close to
the manifolds, but are  unconstrained in other regions in the solution; 
hence leading to a Casimir-type fluctuation-induced attraction.

\end{multicols}


\begin{references}
\bibitem{Retardation}
For liquid Helium films with thickness of more than $100 \AA$, the van der
Waals interaction is retarded, and falls off as $1/H^3$, to leading order. 
In this case, the expression for the film thickness is more complicated,
and will not be elaborated on here.
\bibitem{ChargeA}
The calculation of the determinant using the true compresibility profile
for the Casimir geometry has indeed been carried out in Ref.\protect\cite{Attard},
with the result
\begin{eqnarray}	
&&F_{FI}= -k_BT\times {A\over H^2}\times\nonumber\\
&&\left[ {\zeta(3)\over 16\pi}+{\pi \over 4}\left({\pi \over 4}
+{1 \over 2}\right)+{\pi \over 4} \ln(2 z \ell_B  \sigma H)
+O(\xi/H)\right].\nonumber
\end{eqnarray}
\noindent
It is interesting to note that
the correct result 
is considerably stronger than
that of the simple ``Gouy-Chapman"  model (Eq.(\protect\ref{FFIGC})).
\bibitem{correct}
In this formula, the result in 
Ref.\protect\cite{goul} has been corrected by a factor of $1/2$ \protect\cite{GGK}.
\bibitem{h/H}
For two plates separated by $H$, the perturbation series is in
the small parameters $h_1/H$ and $h_2/H$.
For a single plate ($H\to \infty$), the requirement is that the
time and space derivatives of the height funtions should be small.
\bibitem{cutoff}
The divergence of kernels in IIb comes from integrations over
space-time. Given a cut-off $L$ in plate size, and an associated
cutoff $L/c$ in time, the kernels diverge as 
$\exp[(K-2)L/H]/[K(L/H)^3]$, with $K=2 Q H/\pi$. Some 
care is necessary in the order of limits for $(L,H)\to\infty$.
\end{references}
\end{document}